\documentstyle[prl,aps,epsfig]{revtex}

\begin{document}

\sloppy

\title{Observation of breathers in Josephson ladders}

\author{P. Binder, D. Abraimov, and A. V. Ustinov}
\address{Physikalisches Institut III, Universit\"at Erlangen,
E.-Rommel-Str. 1, D-91058 Erlangen, Germany}

\author{S. Flach and Y. Zolotaryuk}
\address{Max-Planck-Institut f\"ur Physik komplexer Systeme,
N\"othnitzer Str. 38, D-01187 Dresden, Germany}

\date{\today}

\wideabs{

\maketitle

\begin{abstract}
  We report on the observation of spatially-localized excitations in a
  ladder of small Josephson junctions. The excitations are whirling
  states which persist under a spatially-homogeneous force due to the
  bias current. These states of the ladder are visualized using a low
  temperature scanning laser microscopy. We also compute breather
  solutions with high accuracy in corresponding model equations. The
  stability analysis of these solutions is used to interpret the
  measured patterns in the $I-V$ characteristics.
\end{abstract}



}

The present decade has been marked by an intense theoretical research
on dynamical localization phenomena in spatially discrete
systems, namely on
discrete breathers (DB). These exact solutions of the underlying
equations of motion are characterized by periodicity in time and
localization in space. Away from the DB center the system approaches a
stable (typically static) equilibrium. (For reviews see
\cite{sfcrw98},\cite{sa97}). These solutions are robust to changes
of the equations of motion, exist in translationally invariant systems
and any lattice dimension. Discrete breathers 
have been discussed in connection with
a variety of physical systems such as large molecules, molecular
crystals
\cite{oe82}, spin lattices \cite{takeno98,ls99}, to name just a few.

For a localized excitation such as a DB, the excitation of plane waves
which might carry the energy away from the DB does not occur due to
the spatial discreteness of the system. The discreteness provides a
cutoff for the wave length of plane waves and thus allows to avoid
resonances of all temporal DB harmonics with the plane waves. The
nonlinearity of the equations of motion is needed to allow for the
tuning of the DB frequency \cite{sfcrw98}.

Though the DB concept was initially developed for conservative
systems, it can be easily extended to dissipative systems
\cite{rsmjas98}. There discrete
breathers become time-periodic spatially localized attractors,
competing with other (perhaps nonlocal) attractors in phase space. The
characteristic property of DBs in dissipative systems is that these
{\em localized excitations} are predicted to persist under the
influence of a {\em spatially homogeneous} driving force. This is due
to the fact, that the driving force compensates the dissipative
losses of the DB.

So far research in this field was predominantly theoretical.
Identifying and analyzing of experimental systems for the direct
observation of DBs thus becomes a very actual and challenging problem. 
Experiments on localization of light propagating in weakly coupled
optical waveguides \cite{hseys98}, low-dimensional crystals
\cite{Swanson99} and anti-ferromagnetic materials \cite{Schwarz99} have
been recently reported.

\begin{figure}[htb]
\vspace{2pt}
\centerline{\epsfig{file=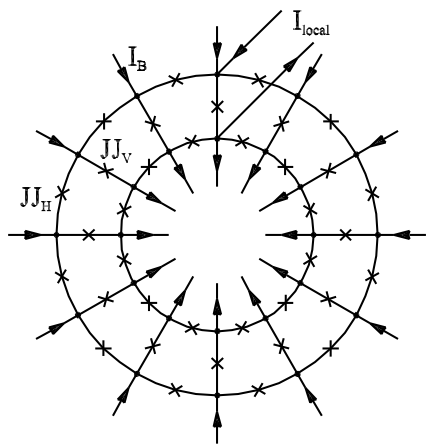,width=2.3in}}
\vspace{2pt}
\caption{Schematic view of an annular ladder. Josephson junctions are
indicated by crosses ($\times$).}
\label{sketch}
\end{figure}

In this work we realize the theoretical proposal \cite{fmmfa96} to
observe DB-like localized excitations in arrays of coupled Josephson
junctions.  A Josephson junction is formed between two superconducting
islands. Each island is characterized by a macroscopic wave function
$\Psi\sim{\rm e}^{{\rm i}\theta}$ of the superconducting state. The
dynamics of the junction is described by the time evolution of the
gauge-invariant phase difference $\varphi=\theta_2 -
\theta_1-\frac{2\pi}{\Phi_0} \int{\bf A}\cdot d{\bf s}$ between
adjacent islands. Here $\Phi_0$ is the magnetic flux quantum and ${\bf
  A}$ is the vector potential of the external magnetic field
(integration goes from one island to the other one).  In the following
we consider zero magnetic fields ${\bf A}={\bf 0}$.  The mechanical
analogue of a biased Josephson junction is a damped pendulum driven by
a constant torque. There are two general states in this system: the
first state corresponds to a stable equilibrium, and the second one
corresponds to a whirling pendulum state. When treated for a chain of
coupled pendula, the DB corresponds to the whirling state of a few
adjacent pendula with all other pendula performing oscillations around
their stable equilibria. In an array of Josephson junctions the nature
of the coupling between the junctions is inductive. A localized
excitation in such a system corresponds to a state where one (or
several) junctions are in the whirling (resistive) state, with all
other junctions performing small forced oscillations around their
stable equilibria.  According to theoretical predictions
\cite{sfms99}, the amplitude of these oscillations should decrease
exponentially with increasing distance from the center of the
excitation.

We have conducted experiments with ladders consisting of
Nb/Al-AlO$_x$/Nb underdamped Josephson tunnel junctions \cite{Hypres}.
We investigated annular ladders (closed in a ring) as well as straight
ladders with open boundaries. The sketch of an annular ladder is given
in Fig.~\ref{sketch}.  Each cell contains 4 small Josephson junctions.
The size of the hole between the superconducting electrodes which form
the cell is about $3\times 3\:\mu$m$^2$.  Here we define {\sl
  vertical} junctions (${\sl JJ_V}$) as those in the direction of the
external bias current, and {\sl horizontal} junctions (${\sl JJ_H}$)
as those transverse to the bias.  Because of fabrication reasons we
made the superconducting electrodes quite broad so that the distance
between two neighboring vertical junctions is 30$\:\mu$m as can be
seen in Fig.~\ref{LTSLMdata}.  The ladder voltage is read across the
vertical junctions.  According to the Josephson relation, a junction
in a whirling state generates a dc voltage
$V=\frac{1}{2\pi}\Phi_0\left<\frac{d\varphi}{dt}\right>$, where
$\left<...\right>$ means the time average. In order to force junctions
into the whirling state we used
two different types of bias. The current $I_{\rm B}$ was uniformly
injected at every node via thin-film resistors.  Another current
$I_{\rm local}$ was applied locally across just one vertical junction.

We studied ladders with different strength of horizontal and vertical
Josephson coupling determined by the junction areas. The ratio of the
junction areas is called the anisotropy factor and is expressed in
terms of the junction critical currents $\eta = I_{\rm cH}/I_{\rm
  cV}$. If this factor is equal to zero, vertical junctions will
  be decoupled and can operate independently one from
  another. Measurements have been performed at $4.2\:$K. The number
of cells $N$ in different ladders varied from 10 to 30. The
discreteness of the ladder is expressed in terms of the parameter
$\beta_{\rm L} = 2\pi L I_{\rm cV}/\Phi_0$, where $L$ is the
self-inductance
of the elementary cell of the ladder. The damping coefficient
$\alpha=\sqrt{\Phi_0/(2\pi I_{\rm c}CR_{\rm N}^2)}$ is the same for
all junctions as their capacitance $C$ and resistance $R_{\rm N}$
scale with the area and $C_{\rm H}/C_{\rm V}=R_{\rm NV}/R_{\rm
  NH}=\eta$. The damping $\alpha$ in the experiment can be controlled
by temperature and its typical values are between $0.1$ and $0.02\:$.

We have measured the dc voltage across various vertical junctions as a
function of the currents $I_{\rm local}$ and $I_{\rm B}$. In order to
generate a localized rotating state in a ladder we started with
applying the local current $I_{\rm local}> 2I_{\rm cH}+I_{\rm cV}$.
This switches one vertical and the nearest horizontal junctions into
the resistive state. After that $I_{\rm local}$ was reduced and,
simultaneously, the homogeneous bias $I_{\rm B}$ was tuned up. In the
final state we kept the bias $I_{\rm B}$ constant and reduced $I_{\rm
  local}$ to zero. Under these conditions, with a {\em
  spatially-homogeneous} bias injection, we observed a {\em
  spatially-localized} rotating state with non-zero dc voltage drops
on just one or a few vertical junctions.

\begin{figure}[htb]
\vspace{2pt}
\centerline{\epsfig{file=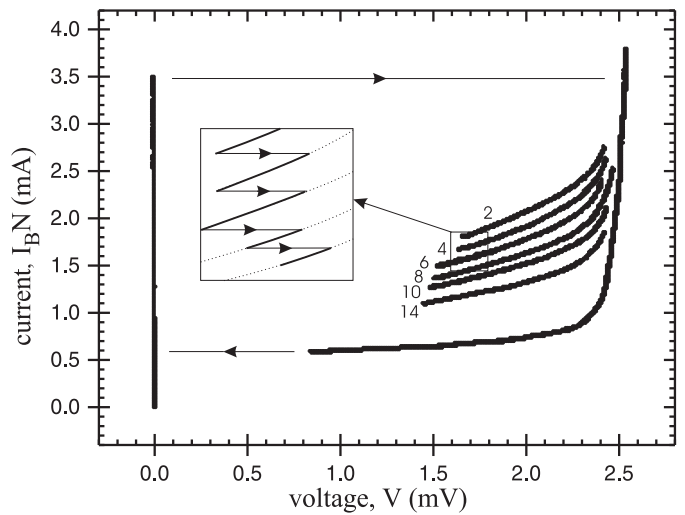,width=3.5in}}
\vspace{5pt}
\caption{
  Current-voltage $I_{\rm B}-V$ plane for an annular ladder with the
  parameters $N=30$, $\eta=0.44$ and $\beta_{\rm L}=2.7$. Digits
  indicate the number of rotating vertical junctions.}
\label{ladder-IVs}
\end{figure}

\begin{figure}[htb]
\vspace{5pt}
\centerline{\epsfig{file=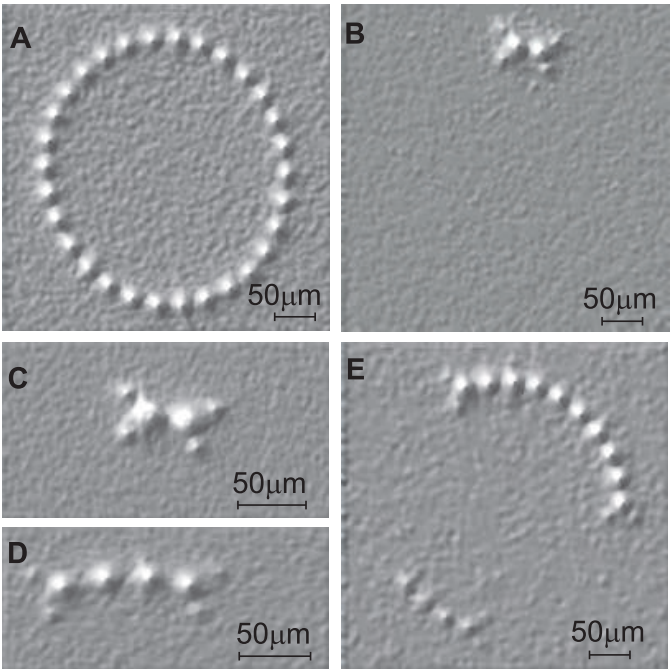,width=3.5in}}
\vspace{5pt}
\caption{
  Whirling states measured in the annular ladder using the low
  temperature scanning laser microscope: (A) spatially-homogeneous
  whirling state, (B)--(E) various localized states corresponding to
  discrete breathers.}
\label{LTSLMdata}
\end{figure}

Various measured states of the annular ladder in the current-voltage
$I_{\rm B}-V$ plane with $I_{\rm local}=0$ are presented in
Fig.~\ref{ladder-IVs}. The voltage $V$ is recorded locally on the same
vertical junction which was initially excited by the local current
injection. The vertical line on the left side corresponds to the
superconducting (static) state. The rightmost (also the bottom) curve
accounts for the spatially-homogeneous whirling state (all vertical
junctions rotate synchronously). Its nonlinear $I_{\rm B}(V)$ shape is
caused by a strong increase of the normal tunneling current at a
voltage of about $2.5\:$mV corresponding to the superconducting energy
gap. The series of branches represent various localized states. These
states differ from each other by the number of rotating vertical
junctions.

In order to visualize various rotating states in our ladders we used
the method of low temperature scanning laser microscopy \cite{LTSLM}.
It is based on the mapping of a sample voltage response as a function
of the position of a focused low-power laser beam on its surface. The
laser beam locally heats the sample and, therefore, introduces an
additional dissipation in the area of few micrometers in diameter.
Such a dissipative spot is scanned over the sample and the voltage
variation at a given bias current is recorded as a function of the
beam coordinate. The resistive junctions of the ladder contribute to
the voltage response while the junctions in the superconducting state
show no response. The power of the laser beam is modulated at a
frequency of several kHz and the sample voltage response is measured
using a lock-in technique.

Several examples of the ladder response are shown in
Fig.~\ref{LTSLMdata} as 2D gray scale maps.  The spatially-homogeneous
whirling state is shown in Fig.~\ref{LTSLMdata}(A).  Here all vertical
junctions are rotating but the horizontal ones are not.
Fig.~\ref{LTSLMdata}(B) corresponds to the uppermost branch of
Fig.~\ref{ladder-IVs}. We observe a localized whirling state expected
for a DB, namely a rotobreather \cite{fmmfa96}. In this case 2
vertical junctions and 4 horizontal junctions of the DB are rotating,
with all others remaining in the superconducting state.  The same
state is shown on an enlarged scale in Fig.~\ref{LTSLMdata}(C).
Fig.~\ref{LTSLMdata}(D) illustrates another rotobreather found for the
next lower branch of Fig.~\ref{ladder-IVs} at which 4 vertical
junctions are in the resistive state. The local current at the
beginning of each experiment is passing through the vertical junction
being the top one on each map. In Fig.~\ref{LTSLMdata}(E), which
accounts for one of the lowest branches, we find an even broader
localized state.  Simultaneously, on the opposite side of the ring we
observe another DB excited spontaneously (without any local current).
An interesting fact is that in experiments with open boundary
ladders (not closed in a ring) we also detected DBs with even or odd
numbers of whirling vertical junctions. 

Various states shown in Fig.~\ref{LTSLMdata} account for different
branches in the $I_{\rm B}-V$ plane in Fig.~\ref{ladder-IVs}. Each
resistive configuration is found to be stable along its particular
branch. On a given branch the damping of the junctions in the rotating
state is compensated by the driving force of the bias current $I_{\rm
  B}$. The transitions between the branches are discontinuous in
voltage. In Fig.~\ref{ladder-IVs}, we see that all branches of
localized states lose their stability at a voltage of about $1.4\:$mV.
Furthermore, as indicated in the inset on Fig.~\ref{ladder-IVs}, a
peculiar switching occurs: upon {\em lowering} the bias current
$I_{\rm B}$ the system switches to a {\em larger} voltage. According
to our laser microscope observations, the lower is the branch in
Fig.~\ref{ladder-IVs}, the larger is the number $M$ of resistive
vertical junctions. The slope of these branches is $dV/dI_{\rm
  B}\approx MR_{\rm NV}/(M+\eta)$, thus the branches become very close
to each other for large $M$. The fact that the voltage at the onset of
instability is independent of the size of the breather, indicates that
the instability is essentially local in space and occurs at the border
between the resistive and nonresistive junctions.

The occurrence of DBs is inherent to our system. We have also found
various
localized states to arise without any local current. Namely, when
biasing
the ladder by the homogeneous current $I_{\rm B}$ slightly below
$NI_{\rm cV}$, we sometimes observed the system switching to a
spatially-inhomogeneous state, similar to that shown in
Fig.~\ref{LTSLMdata}(E).

To interpret the experimental observations, we analyze the equations
of motion for our ladders (see \cite{sfms99} for details). 
Denote by $\varphi^v_l, \varphi^h_l,
\tilde{\varphi}^h_l$ the phase differences across the $l$th vertical
junction and its right upper and lower horizontal neighbors. Using
$\nabla \varphi_l = \varphi_{l+1} - \varphi_l$ and $\Delta \varphi_l =
\varphi_{l+1} + \varphi_{l-1} - 2\varphi_l$, the Josephson equations
yield

\begin{eqnarray}
\ddot{\varphi}^v_l + \alpha \dot{\varphi}^v_l + \sin \varphi^v_l =
\gamma
-\frac{1}{\beta_{\rm L}}(-\Delta \varphi^v_l + \nabla
\tilde{\varphi}^h_{l-1}
- \nabla \varphi^h_{l-1}) 
\label{1-1} \\
\ddot{\varphi}^h_l + \alpha \dot{\varphi}^h_l + \sin \varphi^h_l =
-\frac{1}{\eta \beta_{\rm L}}(\varphi^h_l - \tilde{\varphi}^h_l +
\nabla \varphi^v_l)
\label{1-2} \\
\ddot{\tilde{\varphi}}^h_l + \alpha \dot{\tilde{\varphi}}^h_l + 
\sin \tilde{\varphi}^h_l =
\frac{1}{\eta \beta_{\rm L}}(\varphi^h_l - \tilde{\varphi}^h_l +
\nabla \varphi^v_l)
\label{1-3}
\end{eqnarray}

\begin{figure}[htb]
\vspace{5pt}
\centerline{\epsfig{file=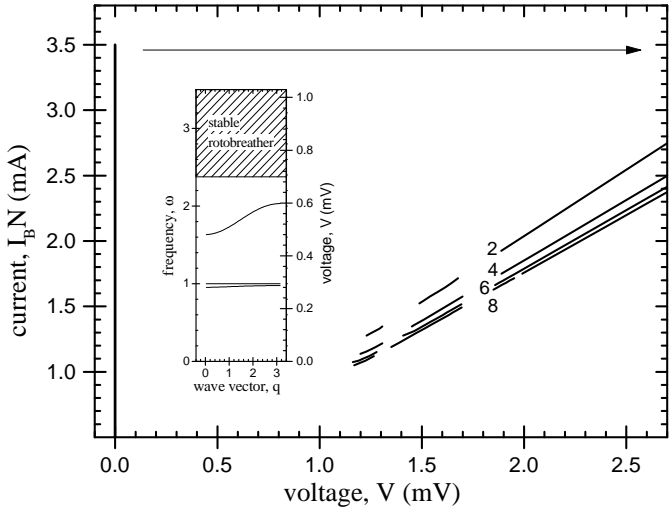,width=3.5in}}
\vspace{5pt}
\caption{
  Current-voltage characteristics for 
  numerically obtained breather states with 2, 4, 6, and 8 whirling
  vertical junctions ($\alpha=0.07$).
  The inset shows the dispersion relation of an
  annular ladder with $\gamma=0.3$ and $N=30$, $\eta=0.44$,
  $\beta_{\rm L}=2.7$.}
\label{dispersion}
\end{figure}

Here $\gamma = I_{\rm B} / (NI_{\rm cV})$.  First, we compute the
dispersion relation for Josephson plasmons $\varphi \propto {\rm
  e}^{{\rm i}(ql-\omega t)}$ at $\alpha=0$ in the ground state (no
resistive junctions). We obtain three branches: one degenerated with
$\omega=1$ (horizontal junctions excited in phase), the second one
below $\omega=1$ with weak dispersion (mainly vertical junctions
excited) and finally the third branch with the strongest dispersion
above the first two branches (mainly horizontal junctions excited out
of phase), cf.\ the inset in Fig.~\ref{dispersion}. The region of
experimentally observed breather stability is also shown. Note, that
for DBs with symmetry between the upper and lower horizontal
junctions the voltage drop on the horizontal junctions is half the
drop across the vertical ones. This causes the characteristic
frequency of the DB to be two times smaller than the value
expected from the measured voltage drop on the vertical
junctions\cite{sfms99}.

In order to compare experimental results of Fig.~\ref{ladder-IVs} to
the model given by Eqs.~(\ref{1-1})-(\ref{1-3}) we have integrated the
latter equations numerically.  We also find localized breather
solutions, in
particular solutions similar to the ones reported in previous
numerical studies \cite{sfms99}.  These solutions are generated with
using initial conditions when M vertical junctions of the resistive
cluster (cf.  Fig.~\ref{LTSLMdata}) have $\varphi=0$ and
$\dot{\varphi}=2V$ and the horizontal junctions adjacent to the
vertical resistive cluster have $\varphi=0$ and $\dot{\varphi}=V$,
while all other phase space variables are set to zero.  The obtained
current-voltage characteristics are shown in Fig.~\ref{dispersion}.
Note that the superconducting gap structure and the
nonlinearity of slopes are not reproduced in the simulations, as we
use
a voltage independent dissipation constant $\alpha$ in
of (\ref{1-1})-(\ref{1-3}).  We find several instability windows of DB
solutions, separating stable parts of the current-voltage
characteristics. 

In addition to direct numerical calculation of $I_{\rm B}-V$ curves,
we have also computed numerically exact breather solutions of
(\ref{1-1})-(\ref{1-3}) by using a generalized Newton map
\cite{sfcrw98}. We have studied the linear stability of the obtained
breather \cite{sfcrw98} by solving the associated eigenvalue problem.
The spatial profile of the eigenmode which drives the instability 
(associated with the edges of
the instability windows in Fig.4) is localized on
the breather, more precisely on the edges of the resistive domain.
This is in accord with the experimental observation (Fig.3) where
several independent breathers can be excited in the system.

The DB states turn to be either invariant under ${\varphi}^h_l
\leftrightarrow {\tilde{\varphi}}^h_l$ transformation or not. Both
such solutions have been obtained numerically. To understand this, we
consider the equations of motion (\ref{1-1})-(\ref{1-3}) in the limit
$\eta\rightarrow 0$ and look for time-periodic localized solutions. In
this limit the brackets on right hand side of (\ref{1-2}) and
(\ref{1-3}) vanish, and vertical junctions decouple from each other.
Let us then choose one vertical junction with $l=0$ to be in a
resistive state and all the others to be in the superconducting state.
To satisfy periodicity in the horizontal junction dynamics we arrive
at the condition
\begin{eqnarray}
\varphi^h_0 = \frac{1}{k}\varphi^v_0 \;\;,\;\;
\tilde{\varphi}^h_0 = -\frac{k-1}{k}\varphi^v_0
\;\;{\rm or}  \nonumber \\
\varphi^h_{0} = \frac{k-1}{k}\varphi^v_0 \;\;,\;\;
\tilde{\varphi}^h_{0} = -\frac{1}{k}\varphi^v_0
\label{2}
\end{eqnarray}
and a similar set of choices for
$\varphi^h_{-1},\tilde{\varphi}^h_{-1}$, with all other horizontal
phase differences set to zero. Here $k$ is an arbitrary positive
integer. Continuation to nonzero $\eta$ values should be possible
\cite{rsmjas98}. The  symmetric DBs in Fig.3
correspond to $k=2$.  The mentioned asymmetric DBs correspond to
$k=1$. The
current-voltage characteristics for asymmetric $k=1$ DBs show a
different behavior from that discussed above. These DBs are stable
down to very small current values, and simply disappear upon further
lowering of the current, so that the system switches from a state with
finite voltage drop to a pure superconducting state with zero voltage
drop.

The observed DB states are clearly different for the well known row
switching effect in 2D Josephson junction arrays. The DBs demonstrate
localization transverse to the bias current (driving force), whereas
the switched states of {\em non-interacting junction rows} are
localized along the current. At the same time, DB states inherent to
Josephson ladders are closely linked to the recently discovered
meandering effect in 2D arrays \cite{Our-meandering:PRL-99}.

In summary, we have experimentally detected various types of
rotobreathers in Josephson ladders and visualized them with the help
of laser microscopy \cite{etjjmtpo99}. Our experiments show that DBs
in Josephson ladders may occupy several lattice sites and that the
number of occupied sites may increase at specific instability points.
The possibility of exciting DBs spontaneously, without using any local
force, demonstrates their inherent character. The observed DBs are
stable in a wide frequency range. Numerical calculations confirm the
reported interpretation and allow for a detailed study of the observed
instabilities.


\begin{thebibliography}{99}

\bibitem{sfcrw98}
S.~Flach and C.~R. Willis, Phys. Rep. {\bf 295}, 181 (1998).

\bibitem{sa97}
S.~Aubry, Physica D {\bf 103}, 201 (1997).

\bibitem{oe82} A.~A. Ovchinnikov and H.~S. Erikhman, Uspekhi Fiz. Nauk
  (Russian) {\bf 138}, 289 (1982).

\bibitem{takeno98}
S.~Takeno, M. Kubota, and K. Kawasaki, Physica D {\bf 113}, 366
(1998).

\bibitem{ls99}
R.~Lai and A.~J. Sievers, Physics Reports, in print (1999).

\bibitem{rsmjas98}
R.~S.~ MacKay and J.~A. Sepulchre, Physica D {\bf 119}, 148 (1998).

\bibitem{hseys98}
H.~S. Eisenberg, Y.~Silberberg, R.~Morandotti, A.~R. Boyd, and
J.~S. Aitchison, Phys. Rev. Lett. {\bf 81}, 3383 (1998).

\bibitem{Swanson99} B. I. Swanson, J. A. Brozik, S. P. Love, G. F.
Strouse, A. P. Shreve, A. R. Bishop, W.-Z. Wang, and M. I. Salkola,
Phys. Rev. Lett. {\bf 82}, 3288 (1999).

\bibitem{Schwarz99} U. T. Schwarz, L. Q. English, and A. J. Sievers,
Phys.  Rev. Lett. {\bf 83}, 223 (1999).

\bibitem{fmmfa96} L.~M. Floria, J.~L. Marin, P.~J. Martinez, F.~Falo,
  and S.~Aubry, Europhys. Lett. {\bf 36}, 539 (1996).

\bibitem{sfms99}
S.~Flach and M.~Spicci, J. Phys.: Condens. Matter {\bf 11}, 321
(1999).

\bibitem{Hypres} HYPRES Inc., Elmsford, NY 10523.

\bibitem{LTSLM}A. G. Sivakov, A.~P. Zhuravel', O.~G. Turutanov, I.~M.
  Dmitrenko, Appl. Surf. Sci. {\bf 106}, 390 (1996).

\bibitem{Our-meandering:PRL-99} D. Abraimov, P. Caputo, G. Filatrella,
M. V. Fistul, G. Yu. Logvenov, and A. V. Ustinov. Broken symmetry of
row switching in 2D Josephson junction arrays. Manuscript LD7783,
accepted for publication in Phys. Rev. Lett. (1999).
  
\bibitem{etjjmtpo99} After this work was completed we became
  aware of the experiment by E.~Trias, J.~J.~Mazo, and
  T.~P.~Orlando (cond-mat/9904144) who detected 1- and 2-site
  roto-breathers by using several voltage probes across a ladder. We
  note that, in contrast to their standard technique limited by a
  number of voltage probes, our method allows for {\em direct visual
    observation} of any multi-site breathers.

\end{thebibliography}
\end{document}